\documentclass[11pt,preprint,superscriptaddress,aps,showkeys,nofootinbib]{revtex4}
\usepackage{amssymb,amsmath,amsfonts}
\usepackage{graphicx}
\usepackage{eepic,epsfig}
\usepackage{indentfirst}
\usepackage[utf8]{inputenc}
\usepackage[T1]{fontenc}
\usepackage{bm} 

1.60cm \makeatletter \@addtoreset{equation}{section}

\makeatother

\begin{document}
	\title{Electromagnetic Casimir effect in a Lorentz symmetry violation model}
	\author{D\^eivid R. da Silva}
	\affiliation{Departamento de F\'{\i}sica, Universidade Federal da Para\'{\i}ba\\
		Caixa Postal 5008, 58051-970, Jo\~ao Pessoa, Para\'{\i}ba, Brazil}
	\email{drds.sensei@gmail.com, emello@fisica.ufpb.br}
	\author{E. R. Bezerra de Mello}
	\affiliation{Departamento de F\'{\i}sica, Universidade Federal da Para\'{\i}ba\\
		Caixa Postal 5008, 58051-970, Jo\~ao Pessoa, Para\'{\i}ba, Brazil}
	\email{drds.sensei@gmail.com, emello@fisica.ufpb.br}

	
	\begin{abstract}
	In this paper, we study the electromagnetic Casimir effects in the context of Lorentz symmetry violations. Two distinct approaches are considered: the first one is based on Horava-Lifshitz methodology, which explicitly presents a space–time anisotropy, while the second is a model that includes higher-derivatives in the field strength tensor and a preferential direction in the space-time. We assume that the electromagnetic field obeys the standard boundary conditions on two large parallel plates. Our main objectives are to investigate how the Casimir energy and pressure are modified in both Lorentz violation scenarios. 
	\end{abstract}
	
	\maketitle
	
	\section{Introduction}
	
The Casimir effect, theoretically proposed by H. B Casimir in 1948 \cite{casimir1948attraction} and experimentally confirmed ten years later by 	M. J. Sparnnaay \cite{sparnaay1958measurements},\footnote{In the 90s, experiments have confirmed the Casimir effect with high degree of accuracy \cite{Lamoureux1997}.} is one of the most important microscopic manifestations of the existence of quantum vacuum fluctuations. Under a classical field point of view, it is not expected any interactions between two uncharged conductive plates in a vacuum; however in the quantum electrodynamics the plates do affect the virtual photons which constitute the field generating a net force. Although the Casimir effect can be expressed in terms of virtual particles interacting with the objects, it is best described and more easily calculated in terms of the zero-point energy of a quantized field in the intervening space between the objects. In his original work, Casimir predicted that due to quantum fluctuations of the electromagnetic field, two parallel flat neutral (grounded) plates attract each other with a force in natural unity given by:
\begin{equation}
F = - \frac{A\pi^2}{240 a^4}\   ,
\end{equation}
being $A$ the area of plates and $a$ is the distance between them. The Casimir effect is traditionally studied by changing up the idealized effects of borders by boundary conditions.
	
In general, we can define the Casimir effect as being a stress (force per unit area) when boundary conditions are imposed on quantum fields. These boundaries can be material means, interfaces between two phases of the vacuum, or even, space-time topologies. Regarding to electromagnetic fields, it is imposed that the parallel component of the electric field, and the perpendicular component of the magnetic field vanish on the plates.
	
Since Quantum Field Theory is based on Relativity, the Lorentz symmetry is preserved. However, other theories propose models where the Lorentz symmetry is violated, as consequence, the space-time anisotropy modifies the Hamiltonian operator spectrum.

In the quantum gravity, Ho\u rava-Lifshitz (HL) theory is a approach where the Lorentz symmetry is not preserved. In this theory, there exist an anisotropy between space and time, that happens due to different properties of scales in which coordinates space and time are set. In this way, the theory is invariant under the rescaling $x\to bx$, $t\to b^{\xi}t$, being $\xi$ a number named critical exponent \cite{hovrava2009quantum}. 

In addition, the violation of Lorentz symmetry has been questioned, in the theoretical and experimental context. In 1989, V. A. Kostelecky and S. Samuel \cite{kostelecky1989spontaneous} described a mechanism in string theory that allows the violation of Lorentz symmetry at the Planck scale. By this mechanism a non-vanishing expectation value of some vector and tensor explicitly imply preferential direction in space-time, producing spontaneous violation of the Lorentz symmetry.	

If there exist a violation of the Lorentz symmetry at the Planck energy scale in a more fundamental theory, the effects of this breakdown must manifest itself in other energy scales. Other mechanisms of violation of Lorentz symmetry are also possible, like space-time non-commutativity \cite{carroll2001noncommutative, anisimov2002remarks, carlson2001bounding, hewett2001signals}, variation of coupling constants \cite{kostelecky2003spacetime, anchordoqui2003time, bertolami1997lorentz}  and modifications of quantum gravity \cite{alfaro2000quantum, alfaro2002loop}.
	
The violation of the Lorentz symmetry became of great experimental interest. Modern experiments have shown the high accuracy of the results obtained through QFT. Under this perspective, the Casimir effect experiments may be useful  in search of vestiges left by the Lorentz symmetry breaking. Withing  this objective, theoretical analysis have been formulated to investigate the consequences of Lorentz violation on the Casimir energy. 

The study of Casimir energy and pressure associated with massless bosonic fields confined between two large and parallel plates in the context of HL theory, has been developed in \cite{ferrari2013hovrava, ulion2015casimir}. As to massive scalar field, the Casimir energy was investigated in \cite{maluf2019casimir}. The analysis of Casimir energy associated to massless fermionic field in HL context was developed in \cite{deivid2019fermionic}. In the context of Lorentz violation CPT-even aether-like, the Casimir energy and pressure has been analyzed, for massive scalar and fermionic field,  in \cite{cruz2017casimir} and \cite{cruz2019fermionic}, respectively. The thermal effects on the Casimir energy associated to massive scalar field, was developed in \cite{cruz2018thermal}.

In this present paper we want to continue in this line of investigation; however analyzing the Casimir energy and pressure associate to electromagnetic field confined between two large and parallel plates in the context of HL and aether-like formalism.

This paper is organized as follows. In Section \ref{Sec2} we introduce the first  model that we want to investigate. This model is a variant of the standard HL model for electrodynamics. In this model we include, besides the standard Maxwell Lagrangian density, an extra lagrangian that explicitly present an anisotropy  between the space and time. In this new term, that only contains contribution of magnetic fields, there appears high order space derivative. We construct the modified field equation and present by adopting Coulomb gauge, the dispersion relation. Also we present explicitly the boundary condition obeyed by the electric and magnetic field on the borders. In Section \ref{Sec3} we present the Hamiltonian operator for this system, and explicitly calculate, by using the Abel-Plana summation formula, a general expression for the Casimir energy. Because this result is not very enlightening, we present an approximated expression and compute the leading order correction term to the Casimir energy. In Section \ref{Sec4}, we calculated the Casimir energy and pressure in a theory that is composed by the Maxwell Lagrangian density and extra  terms that contain higher-derivative of the electromagnetic field tensor through the d'Alembertian operator and second order differential operator $(u\cdot\partial)^2$, being $u^\mu$ an arbitrary constant unity vector, defining a preferential direction in the space-time. By using the generalized dispersion relation, we were able to compute the leading order corrections to the Casimir energy considering different directions for the unity vector. In  Section \ref{concl} we summarize the results obtained in the paper. Here, we assume  $\hbar=c=1$, and the metric signature will be taken as $(+,-,-,-)$.

\section{The Photon Free Field In A Theory With Anisotropy Between Space and Time}
\label{Sec2}	

As we know, the standard Maxwell Lagrangian density governs the dynamics of the electromagnetic field. This Lagrangian is invariant under gauge and Lorentz transformations \cite{griffithsELECT}. In this section we want to analyze the implications on the electrodynamics that smoothly violates Lorentz symmetry, but preserves the gauge invariance.
	
An approach based in Horava-Lifshitz (H-L) theory inspired Farias at all  \cite{Petrov2012}, to propose a Lagrangian to the scalar QED where time and space have different weights. In this case we say that the Lorentz symmetry is broken in a hard way. In this section, we will consider the Lagrangian given bellow to analyze the radiation field in free space. This Lagrangian corresponds to a modification on the one proposed in \cite{Petrov2012}:
\begin{equation}
\mathcal{L} = -\frac{1}{4}F_{\mu \nu}F^{\mu \nu} + \frac{l^{2\varepsilon}}{4} (-1)^{\varepsilon - 1}  F_{ij} \left( \nabla^2 \right)^\varepsilon F^{ij} \ ,
\label{Eq:lagrangian}
\end{equation}
being $F_{\mu \nu} \equiv \partial_\mu A_\nu - \partial_\nu A_\mu$ the  electromagnetic field tensor and $A^\mu \equiv (\phi,\vec{A})$ the four-vector potential. The constant $l$, introduced in the second term of the above expression has dimension of length and the parameters $\varepsilon$ is called critical exponent. Of course \eqref{Eq:lagrangian} recovers the standard Maxwell Lagrangian when we take $l=0$. Because Maxwell equations explain in a very good approximation all the radiation problems, if really there exist a Lorentz violation the parameter $l$ must be quite small.  So, our main objective is to determine how the Casimir energy, $E_C$, depends on the parameter $l$ and $\varepsilon$, and in this way open the possibility of experimental  measurements of $E_C$ impose some restrictions on their values.
	
The Euler-Lagrange equation associated with \eqref{Eq:lagrangian}, is given by
\begin{equation}
\partial_\lambda \left( \frac{\partial \mathcal{L}}{ \partial \left( \partial_\lambda A_\sigma \right) } \right) + \left( \nabla^2 \right)^\varepsilon \partial_\lambda \left\{ \frac{\partial \mathcal{L}}{\partial \left[ \left( \nabla^2 \right)^\varepsilon \partial_\lambda A_\sigma \right]} \right\} = 0 \  ,
\end{equation}
whose explicit differential equations in Coulomb gauge are:
\begin{gather}
\nabla^2 \phi = 0; \label{eq:EDP_phi} \\
\Box \vec{A}+{\vec{\nabla}}\left(\frac{\partial \phi}{\partial t}\right)+ (-1)^{\varepsilon - 1}l^{2\varepsilon} \left( \nabla^2 \right)^{\varepsilon + 1} \vec{A} = \vec{0} \  . \label{eq:EDP_A} 
\end{gather}

\begin{figure}[h]
\centering
\includegraphics[width=0.5\linewidth]{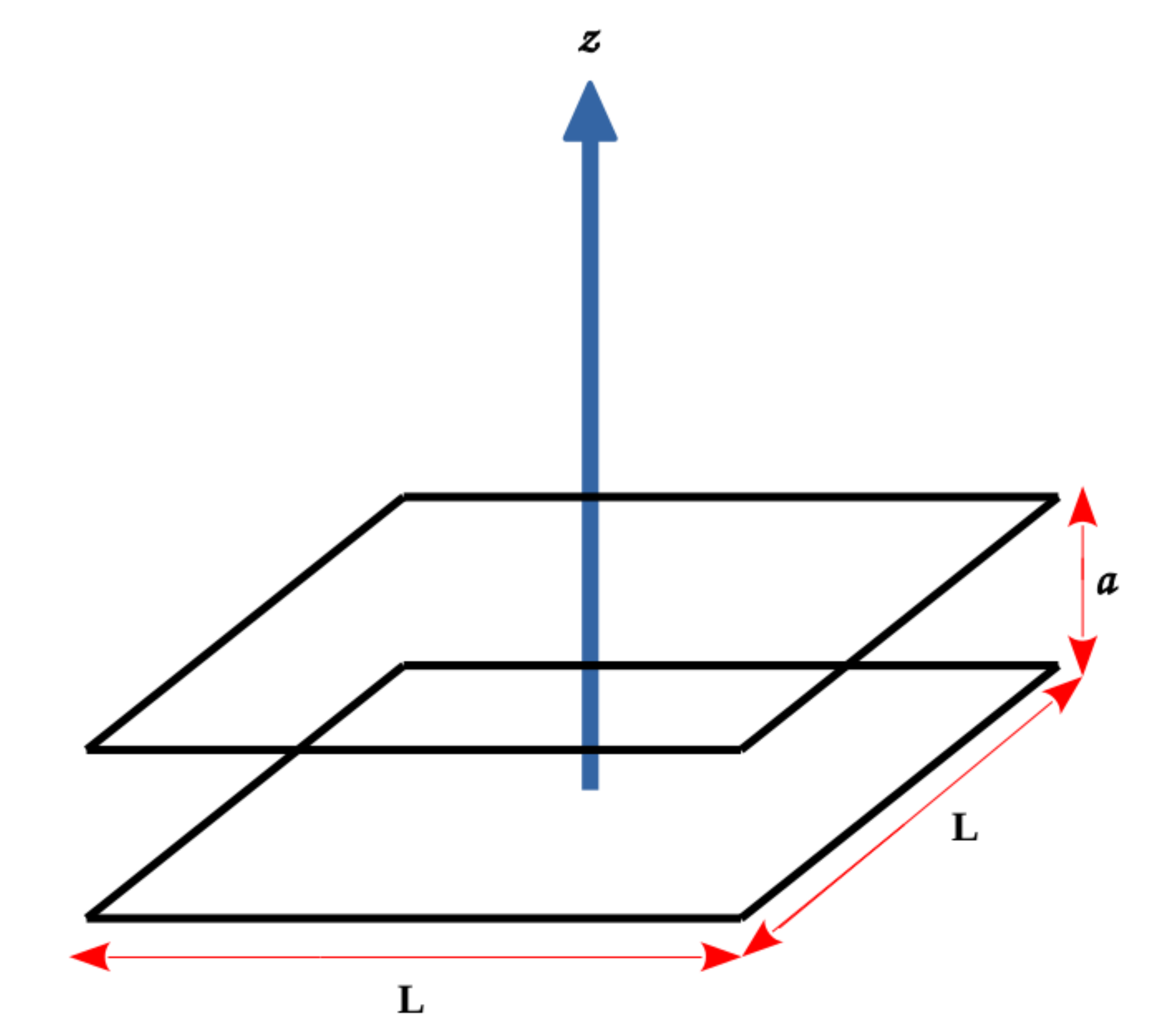}
\caption{Two uncharged parallel plates of areas $L^2$ orthogonal to the $z-$axis, separated by a distance $a$.}
\label{fig:plates}
\end{figure}

Because the main objective of this paper is to investigate the effect of the Lorentz violation on the Casimir energy associated with electromagnetic quantum field confined between two uncharged parallel plates with areas $L^2$, separated a distance $a$, and considering that $L >> a$, (see Figure \ref{fig:plates}), we shall adopt the following boundary conditions \cite{casimir1948attraction, abelplana, itzykson}: 
\begin{gather}
\hat{z} \times \vec{E} \left|_{plates} \right. =  \vec{0}; \label{eq:CC_E}\\
\hat{z} \cdot \vec{B} \left|_{plates} \right. = 0 \  . \label{eq:CC_B}
\end{gather}

Assuming the gauge $\phi=0$, the solution of \eqref{eq:EDP_A} can be obtained in terms of the general expression below for the vector potential:
\begin{equation}
\vec{A} = e^{i(k_x x + k_y y)}\left( \vec{\alpha}e^{ik_z z} + \vec{\beta}e^{-ik_z z}\right)e^{-i\omega t} \  ,
\label{eq:test_solution}
\end{equation}
being $(k_x, \ k_y, \ k_z)$ the three components of the momentum of the photon, and $\omega$ its energy. The two constant vectors $\vec{\alpha}$ and $\vec{\beta}$ are determined by imposing the boundary conditions \eqref{eq:CC_E} and \eqref{eq:CC_B}; however, we can promptly to find the dispersion relation by substituting the general solution (\ref{eq:test_solution}) into \eqref{eq:EDP_A}: 
	\begin{equation}
	\omega^2 = \vec{k}^2 + l^{2\varepsilon} (\vec{k}^2)^{\varepsilon + 1} \  .
	\label{eq:omega}
	\end{equation}
	
This dispersion relation show us that the first term on right hand side corresponds to the energy of a photon in standard electrodynamics \cite{mandlshaw}. The second term is the correction provided by the Lorentz violation, i.e., the photon energy suffers a small perturbation. This interpretation is only possible if we consider that $l^{2\varepsilon} |\vec{k}|^{2\varepsilon} << 1$.
	
The field equations that we found are valid for Coulomb gauge. In this sense we have
	\begin{equation}
	\vec{\nabla} \cdot \vec{A} = 0 \Rightarrow (\vec{k} \cdot \vec{\alpha})e^{ik_z z} + ( \vec{k}^{||} \cdot \vec{\beta}^{||} - k_z\beta_z)e^{-ik_z z} = 0 \  ,
	\end{equation}
where we are using the following notation: $\vec{k}^{||} \equiv k_x \hat{x} + k_y \hat{y}$.

So, after some intermediate steps, we found that the following values for $ k_z $ are acceptable:
\begin{equation}
k_z = \frac{n\pi}{a}, \quad {\rm with} \quad n = 0,1,2, \cdots
\label{eq:k_z}
\end{equation}
Moreover, the solutions compatible with the already mentioned boundary conditions present two possible polarization for the vector potential. The solutions are:
\begin{equation}
\vec{A}_{\vec{k}^{||},n,1} = \frac{1}{2\pi \sqrt{a\omega_{\vec{k}}}} e^{i\vec{k}^{||} \cdot \vec{x}} \sin \left( \frac{n\pi z}{a} \right) \left( \hat{k}^{||} \times \hat{z} \right) e^{-i\omega_{\vec{k}}t};
\label{eq:A_1}
\end{equation}
\begin{equation}
\vec{A}_{\vec{k}^{||},n,2} = \left\{ \begin{array}{c}
\frac{1}{2\pi \sqrt{a\omega_{\vec{k}}}} e^{i\vec{k}^{||} \cdot \vec{x}} \left[ i\frac{n \pi}{a \omega_{\vec{k}}} \sin\left( \frac{n \pi z}{a} \right) \hat{k}^{||} - \frac{|\vec{k}^{||}|}{\omega_{\vec{k}}} \cos \left( \frac{n \pi z}{a} \right) \hat{z}   \right] e^{-i\omega_{\vec{k}}t} \quad {\rm for} \quad n\neq 0 \\ 
- \frac{|\vec{k}^{||}|}{2\pi \omega_{\vec{k}} \sqrt{2a\omega_{\vec{k}}}} e^{i\vec{k}^{||} \cdot \vec{x}} e^{-i\omega_{\vec{k}}t} \hat{z} \quad {\rm for} \quad n=0 \  .
\end{array} \right.
\label{eq:A_2}
\end{equation}

	The above solutions are normalized according to
\begin{equation}
\int_{-\infty}^{\infty}dx \int_{-\infty}^{\infty}dy \int_{0}^{a}dz \vec{A}_{\vec{k}^{||},n,\lambda}^{*} \cdot  \vec{A}_{\vec{q}^{||},m,\sigma} = \frac{1}{2\omega_{\vec{k}}}\delta^{(2)}\left(\vec{k}^{||} - \vec{q}^{||} \right) \delta_{nm} \delta_{\lambda \sigma} \  .
\label{eq:normalization}
\end{equation}

Observing the structure of the equations (\ref{eq:A_1}) and (\ref{eq:A_2}), we can see that when the component of moment in normal direction to the plates vanishes, there is only one polarization. Another important feature of this solution is that they have the same structure as in the standard case (see \cite{FARINA2006,barton1970quantum,lang1998casimir}).

Now we are in position to write the vector potential operator, by adopting the Fourier series bellow and promoting the corresponding coefficients to an operators:
	\begin{equation}
	\hat{\vec{A}} = \int_{-\infty}^{\infty} dk_x \int_{-\infty}^{\infty} dk_y \sum_{n=0}^{\infty} \, \sum_{\substack{\lambda = 1,2 \, (n \neq 0) \\ \lambda = 2 \, (n=0)}} \, \left[ \hat{a}_{\vec{k}^{||},n,\lambda} \vec{A}_{\vec{k}^{||},n,\lambda} + \hat{a}^{\dagger}_{\vec{k}^{||},n,\lambda} \vec{A}_{\vec{k}^{||},n,\lambda}^{*} \right] \  ,
	\label{eq:linear_combination}
	\end{equation}
being $\hat{a}_{\vec{k}^{||},n,\lambda}$ and $\hat{a}^{\dagger}_{\vec{k}^{||},n,\lambda}$ correspond to the annihilation and creation operators, respectively, characterized by the set of quantum numbers: $\vec{k} = (\vec{k}^{||}, n\pi /a)$ and polarization $\lambda$. These operators obey the commutation relations for bosons \cite{mandlshaw}:
	\begin{gather}
	\left[ \hat{a}_{\vec{k}^{||},n,\lambda}, \hat{a}^{\dagger}_{\vec{q}^{||},m,\sigma} \right] = \delta^{(2)}\left(\vec{k}^{||} - \vec{q}^{||} \right) \delta_{nm} \delta_{\lambda \sigma} \ ; \label{eq:boson_commutation}\\
	\left[ \hat{a}_{\vec{k}^{||},n,\lambda} \ , \hat{a}_{\vec{q}^{||},m,\sigma} \right] = \left[ \hat{a}^{\dagger}_{\vec{k}^{||},n,\lambda}, \hat{a}^\dagger_{\vec{q}^{||},m,\sigma} \right] =0 \ .
	\end{gather}

\section{Casimir energy}
\label{Sec3}

Our next step is to find the energy in the region between the plates. Using the definition of Hamiltonian density, we obtain for the Lagrangian density  (\ref{Eq:lagrangian}) the following Hamiltonian operator:
	\begin{equation}
	\hat{H} = \int d^3\vec{x} \left\{ \frac{1}{2}\left( \vec{E}^2 + \vec{B}^2 \right) - \frac{l^{2\varepsilon}}{2} (-1)^{\varepsilon-1} \vec{B} \cdot \left[ (\nabla^2)^\varepsilon \vec{B} \right] \right\}.
	\end{equation}
	
This result is interesting because it tells us that the term with higher order derivatives that we have added in the Lagrangian density (\ref{Eq:lagrangian}) presents exclusively magnetic contribution, i.e., the perturbation considered is provided physically by the magnetic field. Applying the equation (\ref{eq:linear_combination}) together with the orthogonality relation (\ref{eq:normalization}) in the above equation, we find the following expression:
	\begin{equation}
	\hat{H} = \frac{1}{2} \int d^2\vec{k}^{||} \sum_{n=0}^{\infty} \, \sum_{\substack{\lambda = 1,2 \, (n \neq 0) \\ \lambda = 2 \, (n=0)}} \, \left[ \hat{a}^{\dagger}_{\vec{k}^{||},n, \lambda} \hat{a}_{\vec{k}^{||},n, \lambda} + \hat{a}_{\vec{k}^{||},n, \lambda} \hat{a}^{\dagger}_{\vec{k}^{||},n, \lambda} \right] \omega_{\vec{k}^{||},n}.
	\end{equation}
Then, the vacuum energy is
	\begin{equation}
	E_0 = \left< 0\left|\hat{H} \right|0 \right> = \frac{L^2}{8\pi^2} \sum_{n=0}^{\infty} \, \sum_{\substack{\lambda = 1,2 \, (n \neq 0) \\ \lambda = 2 \, (n=0)}} \, \omega_{\vec{k}^{||},n}  \  .
	\label{energy}
	\end{equation}	

Using explicitly the dispersion relation (\ref{eq:omega}), and defining a cylindrical coordinate as $|\vec{k}^{||}| \equiv \rho(\cos(\theta),\sin(\theta)) $, we can rewrite the above equation as follows :
	\begin{equation}
	\begin{split}
	E_0 = & \frac{L^2}{4\pi}\int_{0}^{\infty} d\rho \rho \left( \rho^2 + l^{2\varepsilon}\rho^{2\varepsilon+2}\right)^{\frac{1}{2}} \\
	& + \frac{L^2}{2\pi} \int_{0}^{\infty} d\rho \sum_{n=1}^{\infty} \rho \left\{ \rho^2  + \left( \frac{n\pi}{a} \right)^2 + l^{2\varepsilon} \left[ \rho^2 + \left( \frac{n\pi}{a} \right)^2 \right]^{\varepsilon+1} \right\}^{\frac{1}{2}}.
	\end{split}
	\label{eq:vacuum_energy}
	\end{equation}
	
The integral that involves the sum over the quantum number $n$, will be evaluated by using the Abel-Plana summation formula \cite{abelplana}:
	\begin{equation}
	\sum_{n=1}^{\infty} f(n) = -\frac{1}{2}f(0) + \int_{0}^{\infty} f(t)dt + i \int_{0}^{\infty} \frac{f(it)-f(-it)}{e^{2\pi t}-1}dt \ .
	\label{eq:Abel-Plana}
	\end{equation}
For our case $f(n)$ is
	\begin{equation}
	f(n) = \left\{ \rho^2  + \left( \frac{n\pi}{a} \right)^2 + l^{2\varepsilon} \left[ \rho^2 + \left( \frac{n\pi}{a} \right)^2 \right]^{\varepsilon+1} \right\}^{\frac{1}{2}}.
	\end{equation}
	
Consequently the expression for vacuum energy is formally expressed by
	\begin{small}
		\begin{equation}
		\begin{split}
		E_0 =& \frac{L^2a}{2\pi^2} \int_{0}^{\infty} d\rho \rho \int_{0}^{\infty} du \left[ \rho^2 + u^2 + l^{2\varepsilon} \left( \rho^2 + u^2 \right)^{\varepsilon + 1} \right]^{\frac{1}{2}} + \frac{iL^2a}{2\pi^2}\int_{0}^{\infty} d\rho \rho\int_{0}^{\infty} du \\
		&\frac{\left\{ \rho^2  + \left( iu \right)^2 + l^{2\varepsilon} \left[ \rho^2 + \left( iu \right)^2 \right]^{\varepsilon+1} \right\}^{\frac{1}{2}} - \left\{ \rho^2  + \left( -iu \right)^2 + l^{2\varepsilon} \left[ \rho^2 + \left( -iu \right)^2 \right]^{\varepsilon+1} \right\}^{\frac{1}{2}}}{e^{2au}-1} \  ,
		\end{split}
		\end{equation}
	\end{small}
where we have defined a new variable $u = t\pi /a $.
	
The first term on the right hand side of the above equation corresponds to the vacuum energy in absence of plates as we will show in Appendix \ref{appA}. Its value is divergent and it must be removed in a renormalization procedure \cite{mandlshaw,casimir1948attraction}. Then, the Casimir energy corresponds the second term:
	\begin{small}
	\begin{equation}
	E_c = \frac{iL^2a}{2\pi^2} \int_{0}^{\infty} d\rho \rho \int_{0}^{\infty} du \frac{\left\{ \rho^2  + \left( iu \right)^2 + l^{2\varepsilon} \left[ \rho^2 + \left( iu \right)^2 \right]^{\varepsilon+1} \right\}^{\frac{1}{2}} - \left\{ \rho^2  + \left( -iu \right)^2 + l^{2\varepsilon} \left[ \rho^2 + \left( -iu \right)^2 \right]^{\varepsilon+1} \right\}^{\frac{1}{2}}}{e^{2au}-1} \   .
	\label{eq:E_c}
	\end{equation}
	\end{small}
	
The integral above can be calculated exactly. However, the result is not very enlightening. So, we are going to develop an expansion in the parameter associated with the Lorentz violation. First, we will do two changes of variables: $\rho = w/a$ and $u = s/a$. The next step to expand the term inside the integral in powers of $l/a<<1$. Doing this procedure, we can infer how the Casimir energy is affected by the modification in Maxwell Lagrangian (\ref{Eq:lagrangian}):
	\begin{equation}
	\begin{split}
	E_c =& \frac{iL^2}{2\pi^2a^3} \int_{0}^{\infty} dw w \int_{0}^{\infty} ds \frac{[w^2 + (is)^2]^{\frac{1}{2}} - [w^2 + (-is)^2]^{\frac{1}{2}}}{e^{2s}-1}\\
	&+ \frac{iL^2}{4\pi^2 a^3} \left( \frac{l}{a} \right)^{2\varepsilon} \int_{0}^{\infty} dw w \int_{0}^{\infty} ds \frac{[w^2 + (is)^2]^{\varepsilon + \frac{1}{2}} - [w^2 + (-is)^2]^{\varepsilon + \frac{1}{2}}}{e^{2s}-1} + O((l/a)^{4\varepsilon}) \  .
	\end{split}
	\end{equation}
Using the Euler's formula, we find the following expressions:
	\begin{gather}
	[w^2 + (is)^2]^{\frac{1}{2}} - [w^2 + (-is)^2]^{\frac{1}{2}} = 0 \quad \textrm{for} \quad s<w;\\
	[w^2 + (is)^2]^{\varepsilon + \frac{1}{2}} - [w^2 + (-is)^2]^{\varepsilon + \frac{1}{2}} = 0 \quad \textrm{for} \quad s<w;\\
	[w^2 + (is)^2]^{\frac{1}{2}} - [w^2 + (-is)^2]^{\frac{1}{2}} = 2i (s^2 - w^2)^{\frac{1}{2}} \quad \textrm{for} \quad s > w;\\
	[w^2 + (is)^2]^{\varepsilon + \frac{1}{2}} - [w^2 + (-is)^2]^{\varepsilon + \frac{1}{2}} = 2i \cos (\varepsilon \pi) (s^2 - w^2)^{\varepsilon + \frac{1}{2}} \quad \textrm{for} \quad s > w.
	\end{gather}
	
The four above identities applied to the Casimir energy expression result in the following equation:
	\begin{equation}
	\begin{split}
	E_c =& - \frac{L^2}{\pi^2 a^3} \int_{0}^{\infty} dw w \int_{w}^{\infty} ds \frac{(s^2 - w^2)^{\frac{1}{2}}}{e^{2s}-1}  \\
	& - \frac{L^2\cos (\varepsilon \pi)}{2\pi^2 a^3} \left( \frac{l}{a} \right)^{2\varepsilon} \int_{0}^{\infty} dw w \int_{w}^{\infty} ds \frac{(s^2 - w^2)^{\varepsilon + \frac{1}{2}}}{e^{2s}-1} + O((l/a)^{4\varepsilon}).
	\end{split}
	\end{equation}
Making the changing of variable $s = w \tau$, we can rewrite the above equation as follows:
\begin{equation}
\begin{split}
	E_c =& - \frac{L^2}{\pi^2 a^3} \int_{0}^{\infty} dw w^3 \int_{1}^{\infty} d\tau \frac{(\tau^2 - 1)^{\frac{1}{2}}}{e^{2w\tau}-1}  \\
	& - \frac{L^2\cos (\varepsilon \pi)}{2\pi^2 a^3} \left( \frac{l}{a} \right)^{2\varepsilon} \int_{0}^{\infty} dw w^{2\varepsilon + 3} \int_{1}^{\infty} d\tau \frac{(\tau^2 - 1)^{\varepsilon + \frac{1}{2}}}{e^{2w\tau}-1} + O((l/a)^{4\varepsilon}).
\end{split}
\end{equation}
Again, we will make a variable change to rewrite the Casimir energy expression in a more suitable way. Taking $v= 2w \tau$, the above expression becomes:
	\begin{equation}
	\begin{split}
	E_c =& -\frac{L^2}{16\pi^2 a^3} \int_{1}^{\infty} d\tau \frac{(\tau^2 - 1)^{\frac{1}{2}}}{\tau^4} \int_{0}^{\infty} dv \frac{v^3}{e^v-1}\\
		& - \frac{L^2 \cos (\varepsilon \pi)}{\pi^2 2^{2\varepsilon + 5} a^3} \left( \frac{l}{a} \right)^{2\varepsilon} \int_{1}^{\infty} d\tau \frac{(\tau^2 - 1)^{\varepsilon + \frac{1}{2}}}{\tau^{2\varepsilon + 4}} \int_{0}^{\infty} dv \frac{v^{2\varepsilon + 3}}{e^v-1}+  O((l/a)^{4\varepsilon}).
	\end{split}
	\label{eq:energy_with_integrals}
	\end{equation}

Using the following result:
\begin{equation}
\int_{1}^{\infty} d\tau \frac{(\tau^2-1)^{\alpha + \frac{1}{2}}}{\tau^{2\alpha + 4}} = \frac{1}{2\alpha+3} \left. \frac{(\tau^2 - 1)^{\alpha + \frac{3}{2}}}{\tau^{2\alpha+3}} \right|_{\tau=1}^{\infty} = \frac{1}{2\alpha+3},
\end{equation}
we can find the value of two integrals in Equation (\ref{eq:energy_with_integrals}). The two remaining integrals can be evaluated by using \cite{jeffrey2007table}. So, we finally obtain that Casimir energy per unit area is
\begin{equation}
\frac{E_c}{L^2} = - \frac{\pi^2}{720a^3} - \left( \frac{l}{a} \right)^{2\varepsilon}\frac{\cos (\varepsilon \pi) (2\varepsilon+2)! \zeta(2\varepsilon + 4)}{\pi^2 2^{2\varepsilon+5}a^3} + O((l/a)^{4\varepsilon}) \  ,
\label{eq:solution_energy}
\end{equation}
being $\zeta(z)$ the Riemann zeta function. 

The first term on the right hand side of \eqref{eq:solution_energy} corresponds to the standard Casimir energy \cite{casimir1948attraction}, the second term is the first order energy correction induced by Maxwell's Lagrangian modification (\ref{Eq:lagrangian}). This correction is valid for the case where the ration $\frac{l}{a}<<1$. Moreover, the cosine dependence on the critical exponent makes Casimir energy first order correction be positive or negative. 

The Casimir pressure follows directly from the above equation:
\begin{equation}
P_c = - \frac{1}{L^2} \frac{\partial E_c}{\partial a} = - \frac{\pi^2}{240a^4} - \left( \frac{l}{a} \right)^{2\varepsilon} \frac{\cos (\varepsilon \pi) (2\varepsilon+3)! \zeta(2\varepsilon + 4)}{\pi^2 2^{2\varepsilon+5}a^{4}} + O((l/a)^{4\varepsilon}) \  .
\label{eq:P_c}
\end{equation}
	
\section{Casimir effect in a theory with higher-derivative Lorentz-breaking extension of QED}
\label{Sec4}
	
In this section, we will calculate the expression to Casimir energy and pressure in a theory that presents, besides the Maxwell lagrangian, an extra term that contains higher-derivative in the electromagnetic field tensor, and a preferential direction in the space-time, characterized by a direct coupling between a constant unity vector, $u^\mu$, and the derivative of the field strength. The photon field model that we will use is given in \cite{Petrov2019}, where the Lagrangian density is
	\begin{equation}
	\label{Maxweel_Mod}
	\mathcal{L}_{HD}=-\frac{1}{4}F_{\mu\nu}F^{\mu\nu}-\frac{1}{M}\epsilon^{\beta \mu \nu \lambda}u_{\beta}A_{\mu}\left(c_1(u \cdot \partial)^2-c_2u^2 \Box \right) F_{\nu \lambda} \  , 
	\end{equation}
where $c_1$ and $c_2$ are arbitrary constants chosen conveniently to provide a more workable dispersion relation. 	

The first contribution on the RHS of the above expression is the standard Maxwell term for a free photon. The second is the new one. It includes a  preferential direction $u^\mu$, and higher-derivative of the field strength tensor. The tensor $\epsilon^{\beta \mu \nu \lambda}$ is the Levi-Civita symbol, $M$ is scale energy of the order of Planck mass.
	
In the original paper \cite{Petrov2019}, the authors were able to find the dispersion relation. It is: 
	\begin{equation}
	(\omega^2-\vec{k}^2)^2+\frac{16}{M^2}\Big(c_2u^2(\omega^2-\vec{k}^2)-c_1(u_0 \omega-\vec{u}\cdot\vec{k})^2
	\Big)^2\Big(u^2(\omega^2-\vec{k}^2)-(u_0 \omega-\vec{u}\cdot\vec{k})^2
	\Big)=0.
	\label{eq:general_dispersion_relation}
	\end{equation}

Our main objective in this section is to calculate the correction on the Casimir energy and pressure due to Lorentz symmetry violation associated with a specific choice for the constant vector $u^\mu$. For this reason we will consider only two possibilities in this article: $c_1 = 1$, $c_2 = 0$ and $u^2 = 1$; $c_1 = 1$, $c_2 = 0$ and $u^2 = -1$.

\subsection*{Casimir energy when $c_1 = 1$, $c_2 = 0$ and $u^{\mu} = (1,0)$}
	
	In this case, the dispersion relation (\ref{eq:general_dispersion_relation}) becomes
	\begin{equation}
	(\omega^2 - \vec{k}^2)^2 - \frac{16}{M^2}\vec{k}^2 \omega^4 = 0.
	\end{equation}
	
As we can see there appear a small corrections in the  standard Casimir energy. Before to develop explicitly the calculation of the Casimir energy, let us first adopt a new definition for the momentum by $\vec{k} = (\vec{k}^{||}, n\pi/a) \equiv \vec{u}/a$,  and assuming $(Ma)^{-1} << 1$. Under these condition we can rewrite the above equation as follows:
	\begin{equation}
	\omega = \frac{|\vec{u}|}{a} \pm \frac{1}{Ma} \frac{2}{a}|\vec{u}|^2 + \frac{1}{(Ma)^2} \frac{8}{a}|\vec{u}|^3 + O\left(\frac{1}{(Ma)^3}\right).
	\end{equation}
The first term in this equation is the usual photon energy. The following terms correspond the corrections.
	
Using the approximated photon dispersion relation above into Equation (\ref{energy}), we are able to calculate the vacuum energy correction as:
	\begin{equation}
	\begin{split}
	\Delta E_0 = & \pm \frac{1}{Ma}\frac{L^2}{4\pi^2 a^3} \int d^2 \vec{u}^{||}  (\vec{u}^{||})^2 \pm \frac{1}{Ma} \frac{L^2}{2\pi^2 a^3} \int d^2 \vec{u}^{||} \sum_{n=1}^{\infty} \left[ (\vec{u}^{||})^2 + (n\pi)^2 \right] \\
	& + \frac{1}{(Ma)^2} \frac{L^2}{\pi^2a^3} \int d^2\vec{u}^{||}  |\vec{u}^{||}|^3 + \frac{1}{(Ma)^2} \frac{2L^2}{\pi^2 a^3} \int d^2\vec{u}^{||} \sum_{n=1}^{\infty} \left[ (\vec{u}^{||})^2 + (n\pi)^2 \right]^{\frac{3}{2}}\\
	=& \pm I_1 \pm I_2 + I_3 + I_4.
	\end{split}
	\end{equation}
	
As we did in the last section, we will use the Abel-Plana summation formula, Eq.  (\ref{eq:Abel-Plana}), to transform the summations over the quantum number $n$ above in integrals representations. Doing this we observe that $I_2$ produces a term that cancel $I_1$, contains a divergent term and another one that automatically vanishes. As to $I_4$, it produces a term equal to $-I_3$, a divergent term plus a finite one given in the expression below. Removing the divergent results by a renormalization procedure, we transform the vacuum energy correction into Casimir energy correction:
	\begin{equation}
	\Delta E_c = \frac{1}{(Ma)^2} \frac{2iL^2}{\pi^2 a^3} \int d^2\vec{u}^{||} \int_{0}^{\infty} dt \frac{[(\vec{u}^{||})^2 + (it\pi)^2]^{\frac{3}{2}} - [(\vec{u}^{||})^2 + (-it\pi)^2]^{\frac{3}{2}}}{e^{2\pi t} - 1} \cdot
	\end{equation}
	
These integrals are simplify when we make use of polar coordinates: $\vec{u}^{||} = \rho (\cos \theta, \sin \theta )$. Another useful change of variable is $v=t\pi$. The integrals in parameter $v$ does not vanish only when $\rho < v < \infty$. Then, the expression of Casimir energy correction is
	\begin{equation}
	\Delta E_c = \frac{1}{(Ma)^2} \frac{8L^2}{\pi^2 a^3} \int_{0}^{\infty} d\rho \rho \int_{\rho}^{\infty} dv \frac{(v^2 - \rho^2)^{\frac{3}{2}}}{e^{2v}-1}\cdot
	\end{equation}
	
The following expression is obtained doing two changes of variables in the above equation: $v = \rho \tau$ and $w = 2\rho\tau$.
	\begin{equation}
	\label{correction1}
	\Delta E_c = \frac{1}{(Ma)^2} \frac{L^2}{8\pi^2a^3}\int_{1}^{\infty} d\tau \frac{(\tau^2-1)^{\frac{3}{2}}}{\tau^5} \int_{0}^{\infty} dw \frac{w^5}{e^w-1} = \frac{1}{(Ma)^2} \frac{L^2 \pi^5}{336a^3}\cdot
	\end{equation}

Consequently, the Casimir energy up to the first order correction is
	\begin{equation}
	\frac{E_ c}{L^2} = - \frac{\pi^2}{720a^3} + \frac{1}{(Ma)^2} \frac{\pi^5}{336a^3}\cdot
	\label{eq:CE_time_direction}
	\end{equation}
	
The corresponding Casimir pressure reads
	\begin{equation}
	P_c = - \frac{1}{L^2} \frac{\partial E_c}{\partial a} = - \frac{\pi^2}{240a^4} + \frac{1}{(Ma)^2} \frac{5\pi^5}{336a^4}\cdot
	\end{equation}
	
\subsection*{Casimir energy when $c_1 = 1$, $c_2 = 0$ and $u^{\mu} = (0,\hat{u})$}

In this subsection, we will calculate the Casimir energy and pressure associated to a photon with the following dispersion relation:
	\begin{equation}
	(\omega^2 - \vec{k}^2)^2 - \frac{16}{M^2}(\hat{u}\cdot\vec{k})^4 [\omega^2 - \vec{k}^2 + (\vec{u}\cdot\hat{k})^2] = 0.
	\end{equation}
	
Again, we are considering that $\vec{k} = (\vec{k}^{||}, n\pi/a) \equiv \vec{s}/a$ and $Ma <<1$. Under this condition we can expand the above expression in a power series of $1/(Ma)$:
	\begin{equation}
	\omega = \frac{|\vec{s}|}{a} \pm \frac{1}{Ma} \frac{2(\hat{u}\cdot\vec{s})^3}{a|\vec{s}|} + \frac{1}{(Ma)^2} \frac{4(\hat{u}\cdot \vec{s})^4}{a|\vec{s}|} + O\left(\frac{1}{(Ma)^3}\right).
	\end{equation}

The first term on the RHS provides the standard Casimir energy when applied in Equation (\ref{energy}). The remaining terms produce the following  correction in vacuum energy:
	\begin{equation}
	\Delta E_0 = \pm \frac{1}{Ma} \frac{L^2}{2\pi^2a^3} \int d^2\vec{s}^{||} \sum_{n=1}^{\infty} \frac{(n\pi)^3}{\sqrt{(\vec{s}^{||})^2 + (n\pi)^2}} + \frac{1}{(Ma)^2} \frac{L^2}{\pi^2 a^3} \int d^2\vec{s}^{||} \sum_{n=1}^{\infty} \frac{(n\pi)^4}{\sqrt{(\vec{s}^{||})^2 + (n\pi)^2}}\cdot
	\end{equation}
	
We will use the polar coordinates, $\vec{k}^{||} = \rho(\cos \theta, \sin \theta)$, and the Abel-Plana  summation formula (\ref{eq:Abel-Plana}) to develop the above expression. Doing this the two contributions contain divergent results plus extra terms reproduced below. The divergent results can be removed by an appropriated renormalization procedure. In general this methodology provides the Casimir energy. But, this does not happen here. There appear one more divergent term that will be removed latter. Them, the vacuum energy correction is
	\begin{equation}
	\begin{split}
	\Delta E_0 =& \pm \frac{1}{Ma} \frac{i}{\pi^2 a^3} \int_{0}^{\infty} d\rho \rho \int_{0}^{\infty} \frac{dv}{e^{2v} - 1} \left[ \frac{(iv)^3}{\sqrt{\rho^2 + (iv)^2}} - \frac{(-iv)^3}{\sqrt{\rho^2 + (-iv)^2}}\right]\\
	& + \frac{1}{(Ma)^2}\frac{2i}{\pi^2a^3} \frac{i}{\pi^2 a^3} \int_{0}^{\infty} d\rho \rho \int_{0}^{\infty} \frac{dv}{e^{2v} - 1} \left[ \frac{(iv)^4}{\sqrt{\rho^2 + (iv)^2}} - \frac{(-iv)^4}{\sqrt{\rho^2 + (-iv)^2}}\right] \\
	& + \mathrm{divergent \ terms}.
	\end{split}
	\end{equation}
	
To calculate the above expression, we will use the relations bellow\footnote{All these relations were obtained making use of the Euler's formula \cite{hassani2008mathematical}: $e^{i\theta} = \cos \theta + i \sin \theta$.}:
	\begin{gather}
	\sqrt{\rho^2 + (\pm iv)^2} = \sqrt{\rho^2 - v^2} \quad \textrm{if} \quad v < \rho;\\
	\sqrt{\rho^2 + (\pm iv)^2} = \pm i \sqrt{v^2 - \rho^2} \quad \textrm{if} \quad v > \rho;\\
	\frac{(iv)^3}{\sqrt{\rho^2 + (iv)^2}} - \frac{(-iv)^3}{\sqrt{\rho^2 + (-iv)^2}} = - \frac{2iv^3}{\sqrt{\rho^2 - v^2}} \quad \textrm{if} \quad v < \rho;\\
	\frac{(iv)^3}{\sqrt{\rho^2 + (iv)^2}} - \frac{(-iv)^3}{\sqrt{\rho^2 + (-iv)^2}} = 0 \quad \textrm{if} \quad v > \rho;\\
	\frac{(iv)^4}{\sqrt{\rho^2 + (iv)^2}} - \frac{(-iv)^4}{\sqrt{\rho^2 + (-iv)^2}} = 0 \quad \textrm{if} \quad v < \rho;\\
	\frac{(iv)^4}{\sqrt{\rho^2 + (iv)^2}} - \frac{(-iv)^4}{\sqrt{\rho^2 + (-iv)^2}} = - \frac{2iv^4}{\sqrt{v^2 - \rho^2}} \quad \textrm{if} \quad v > \rho.
	\end{gather}
	
Applying the six above identities in vacuum energy correction we get the following equation:
	\begin{equation}
	\begin{split}
	\frac{\Delta E_0}{L ^2} =& \pm \frac{1}{Ma} \frac{2}{\pi^2a^3} \int_{0}^{\infty} d\rho \rho \int_{0}^{\rho} dv \frac{v^3}{(e^{2v} - 1)\sqrt{\rho^2 - v^2}}\\
	& + \frac{1}{(Ma)^2} \frac{4}{\pi^2a^3} \int_{0}^{\infty} d\rho \rho \int_{\rho}^{\infty} dv \frac{v^4}{(e^{2v} - 1)\sqrt{v^2 - \rho^2}} + \mathrm{divergent \ terms}.
	\end{split}
	\end{equation}
	
Using the same change of variables as in the previews cases, we obtain:
	\begin{equation}
	\begin{split}
	\frac{\Delta E_0}{L ^2} =& \pm \frac{1}{Ma} \frac{1}{16\pi^2 a^3} \int_{0}^{\infty}dw \frac{w^4}{e^w - 1} \int_{0}^{1} d\tau \frac{1}{\tau^2 \sqrt{1-\tau^2}}\\
	& + \frac{1}{(Ma)^2} \frac{1}{8\pi^2 a^3} \int_{0}^{\infty}dw \frac{w^5}{e^w - 1} \int_{1}^{\infty} d\tau \frac{1}{\tau^2 \sqrt{\tau^2 - 1}}+ \mathrm{divergent \ terms}.
	\end{split}
	\end{equation}
	
The first integral is divergent, so, only the second integral on the RHS contribute to Casimir energy correction. After some intermediate steps we get:
	\begin{equation}
	\label{correction2}
	\frac{\Delta E_0}{L ^2} = \frac{1}{(Ma)^2} \frac{\pi^4}{63a^3} \Rightarrow \frac{E_c}{L^2} = - \frac{\pi^2}{720a^3} + \frac{1}{(Ma)^2} \frac{\pi^4}{63a^3}\cdot
	\end{equation}
	
	The Casimir pressure follows directly from this result: 
	\begin{equation}
	P_c = - \frac{1}{L^2} \frac{\partial E_c}{\partial a} = - \frac{\pi^2}{240a^4} + \frac{1}{(Ma)^2} \frac{5\pi^4}{63a^4}\cdot
	\end{equation}
	
Before to finish this subsection we have to explain that the calculation developed in this space-like vector case, took into account only  $u^\mu$ along the $z-$direction. For the other two directions no correction to the Casimir energy were fond up to order $1/(Ma)^2$.  
	
\section{Concuding remarks}
\label{concl}
	\appendix
	
In this paper we have analyzed the influence of the Lorentz violation on the Casimir energy associated with quantum vector field confined between two uncharged parallel palates of area $L^2$ separated by a distance $a<<L$. Two different approaches have been considered. In the first one, proposed by Horava, the Lorentz violation is implemented by an anisotropy between space and time. The second is a model that contains higher-derivative in the field strength and also a preferential direction in the space-time. In both cases, we have shown that there appear  small corrections on the Casimir energy and pressure.

In the first model, where the Lorentz violation is implemented by the anisotropy between space and time through the HL approach, the corrections in Casimir energy and  pressure \eqref{eq:solution_energy} and (\ref{eq:P_c}), respectively,  have a cosine dependence, $\cos (\varepsilon \pi)$. So, depending of value of the critical exponent $\varepsilon$, this correction increases or decreases the force between the plates. Another interesting point of this model is that the dependence of the distance between the plate in the correction is higher than the standard case. The standard Casimir pressure is proportional to $1/a^4$, while the correction is proportional to $1/a^{2\varepsilon + 4}$. In the first glance, it appears that for small distance the correction becomes more relevant than the standard case. However this is not true, because this correction in fact depends on the ratio  $(l/a)^{2\varepsilon}$ that is much smaller than unity.

The second model considers two different ingredients in addition to the Maxwell term. These new ingredients are: higher derivative terms in the tensor field strength, and a preferential space-time direction through the presence of an arbitrary four-vector, $u^\mu$. Because the main objective of this paper is to investigate how the Lorentz violation affects the Casimir energy and pressure, we adopt a particular choice for the coefficient in \eqref{Maxweel_Mod}, i.e., we assumed $c_2=0$ and $c_1=1$.  Adopting, separately, that the vector is time-like or space-like, we were able to calculate the corresponding corrections to the Casimir energy, Eq.s \eqref{correction1} and \eqref{correction2}, respectively. These corrections are of order $O(1/(aM)^2)$. As to the situation where the spatial preferential direction is orthogonal to the plates, it produces a correction in the Casimir pressure that is $1.69765$ bigger than the correction with a preferential direction in time. Moreover, the Casimir pressure corrections in these two cases are proportional to $1/a^6$. And, again increases faster than the standard $1/a^4$. So, the only way to make  these results  plausible is considering $Ma >> 1$.

If the Lorentz violation is a reality, it is possible that all results in the standard QFT have small corrections in a scale that we are not able to access yet.

{\bf Acknowledgements.} The authors want to thank A. Y. Petrov for valuable discussions and comments. D.R.S is supported by Coordenação de Aperfeiçoamento de Pessoal de Nível Superior (CAPES) with number registration 88887.462084/2019-00, and E.R.B.M is partially supported by Conselho Nacional de Desenvolvimento Cient\'{\i}fico e Tecnol\'{o}gico - Brasil (CNPq) under grant No 301.783/2019-3. 

\section{The Vacuum Energy Without Plates }
	\label{appA}
	
	In this Appendix we will show the formal  expression of vacuum energy without plates.
	
	Applying the following periodic conditions for the vector potential:
	\begin{equation}
	\vec{A}(0,y,z,t) = \vec{A}(L,y,z,t), \vec{A}(x,0,z,t) = \vec{A}(x,L,z,t) \quad \textrm{and} \quad \vec{A}(x,y,0,t) = \vec{A}(x,y,a,t),
	\end{equation}
	we can express a free field as a Fourier series \cite{mandlshaw}. Considering a large value for $L$ and $a$ the summation under all possible momentum  becomes a integral. So, in absence oft plates, the vacuum energy $E_v$ in the region with volume $L^2a$ is formally given by
	\begin{equation}
	E_v = \frac{L^2a}{(2\pi)^3}\int d^3\vec{k} \omega_{\vec{k}} = \frac{L^2a}{(2\pi)^3}\int d^3\vec{k} \left[ \vec{k}^2 + l^{2\varepsilon} (\vec{k}^2)^{\varepsilon+1} \right]^{\frac{1}{2}} \ .
	\end{equation}
	Using cylindrical coordinates, $\vec{k} = (\rho \sin \theta, \rho \cos \theta, u)$, the above expression is rewritten as follows:
	\begin{equation}
	E_v = \frac{L^2a}{2\pi^2}\int_{0}^{\infty} d\rho \rho \int_{0}^{\infty} du \left[\rho^2 + u^2 + l^{2\varepsilon}\left( \rho^2 + u^2 \right)^{\varepsilon+1} \right]^{\frac{1}{2}}.
	\end{equation}

\end{document}